\documentclass[prl,twocolumn,showpacs,amsmath,bibnotes,superscriptaddress,floatfix]{revtex4}

\usepackage{graphicx}

\newcommand\pictc[5]{\begin{figure}
                       \centerline{
\includegraphics[width=#1\columnwidth,height=0.7\textheight,keepaspectratio]{#3}}
                   \protect\caption{\protect\label{fig:#4} #5}
                    \end{figure}            }
\newcommand\pict[4][1]{\pictc{#1}{!tb}{#2}{#3}{#4}}
\newcommand\rpict[1]{\ref{fig:#1}}

\newcommand\leqt[1]{\protect\label{eq:#1}}
\newcommand\reqtn[1]{\ref{eq:#1}}
\newcommand\reqt[1]{(\reqtn{#1})}

\newcounter{Fig}

\begin{document}
\begin{sloppy}

\title{Coherence controlled soliton interactions}

\author{Ting-Sen Ku}
\author{Ming-Feng Shih}
\affiliation{Department of Physics, National Taiwan
University, Taipei, 106, Taiwan}

\author{Andrey A. Sukhorukov}
\author{Yuri S. Kivshar}
\affiliation{Nonlinear Physics Centre, Research School of Physical
Sciences and Engineering, Australian National University, Canberra
ACT 0200, Australia}
\homepage{http://www.rsphysse.anu.edu.au/nonlinear}

\begin{abstract}
We demonstrate theoretically and subsequently observe in
experiment a novel type of soliton interaction when a pair of
closely spaced spatial optical solitons {\em as a whole} is made
{\em partially incoherent}. We explain how the character of the
soliton interaction can be controlled by the total partial
incoherence, and show a possibility to change the soliton
interaction from attractive to repulsive, or vice versa, near a
certain threshold in the coherence parameter.
\end{abstract}


\maketitle

In recent years, the study of solitons has attracted considerable
interest in nonlinear optics as well as in other areas of physics
such as Bose-–Einstein
condensates~\cite{Kivshar:2003:OpticalSolitons}. Usually, most
features of solitons are associated with their coherent nature,
and the soliton parameters such as the amplitude, phase, and
frequency are well defined, being also responsible for the
strength and character of the soliton interactions in nonlinear
media.

However, the fundamental concept of solitons as fully coherent
objects was extended in 1996 to cover more general classes of
self-trapped beams which are partially coherent, and therefore can
be termed as {\em incoherent
solitons}~\cite{Mitchell:1996-490:PRL}. Somewhat related concepts
have been developed in the theory of self-trapping of
quasi-periodic waves~\cite{Clausen:1999-4740:PRL} and spontaneously generated
temporal solitons in a nonlinear medium with instantaneous
response~\cite{Picozzi:2004-143906:PRL}.

To support incoherent spatial solitons, the nonlinearity must be
noninstantaneous, such that the medium is insensitive to the fast
changing phase of the light beam, but only can respond to the
time-averaged light intensity. Most experiments on the incoherent
spatial solitons have been performed in the biased photorefractive
material, whose nonlinear response is noninstantaneous due to the
fact that its nonlinearity comes from the migration of space
charges~\cite{Mitchell:1996-490:PRL}. In parallel to the
experimental discovery, several different theoretical
approaches~\cite{Christodoulides:1997-646:PRL,
Mitchell:1997-4990:PRL, Hall:2002-35602:PRE} (for an overview of
different methods, see Ref.~\cite{Kivshar:2003:OpticalSolitons})
to describe incoherent spatial solitons have been developed in the
same period of time. One of the theoretical methods, the coherent
density approach~\cite{Christodoulides:1997-646:PRL}, is a
powerful tool for this kind of analysis, and it is used mostly
when the dynamics of solitons is the primary subject of study.

Another phenomenon associated with the properties of optical
solitons that also caught much research attention is their
particle-like interaction~\cite{Stegeman:1999-1518:SCI}. This has
been demonstrated, in particular, with the bright optical spatial
solitons in self-focusing media. The main features and mechanisms
of the interaction of solitons as coherent objects are well
known~\cite{Gordon:1983-596:OL}. If two bright solitons
are mutually coherent, they attract (repel) each other when they
are in-phase ($\pi$ out of phase). With the relative phase between
the interacting solitons being other than zero or $\pi$, there is
energy transfer from one soliton to the other, in addition to the
repulsive or attractive interaction. This energy transfer is most
significant when the soliton relative phase is $\pi/2$. On the
other hand, if the interacting solitons are mutually incoherent or
both solitons are partially coherent (the relative phase between
them varies much faster than the material can respond), the
soliton interaction is always
attractive~\cite{Anderson:1985-2270:PRA}.

In this Letter, we reveal the existence of a new type of soliton
interaction that is observed when two interacting solitons {\em as
a whole} are made partially incoherent. We find that the soliton
interaction dynamics can become dramatically different from the
case when the solitons are mutually coherent. The interaction
strength can be controlled by the total coherence. Most remarkably, the
interaction may change from attractive to repulsive near a certain
threshold in the coherence parameter, or vice versa.

In order to shed light on this novel type of the soliton
interaction and its origin, first we present our numerical
results, then provide a qualitatively physics intuition to explain
why this happens. Finally, we demonstrate experimentally the
coherence controlled soliton interaction in a biased self-focusing
photorefractive crystal.

In this paper, we use the coherence density
approach~\cite{Christodoulides:1997-646:PRL} and perform numerical
simulations of interacting partially coherent solitons in a planar
geometry. Propagation of light beams in a slow Kerr-type nonlinear
medium can be characterized by interaction of many mutually
incoherent components governed by a set of coupled nonlinear
equations,
\begin{equation} \leqt{nls}
  i \frac{\partial E_n}{\partial z}
  +  D \frac{\partial^2 E_n}{\partial x^2}
  + \gamma \sum_m |E_m|^2 E_n
  = 0,
\end{equation}
where $n$ is the number of component, $\sum_n |E_n|^2 =I$ is the
total beam intensity, $x$ and $z$ are the transverse and
propagation coordinates normalized to their characteristic values
$x_0$ and $z_0$, respectively, $D = \lambda z_0 / (4 n_0 x_0^2)$
is the diffraction coefficient, $\lambda$ is the vacuum
wavelength, $n_0$ is the medium refractive index, and $\gamma$ is
the effective normalized nonlinear coefficient.

When at the crystal input the light is generated  by a partially
incoherent source, one can associate different components with the
same coherent field $E_s(x)$, but having different inclination
angles $n \theta_s$:
\begin{equation}
E_n(z=0) = \sqrt{G(n \theta_s) \theta_s} E_s(x) \exp( i n \theta_s
x k_0),
\end{equation}
where $k_0 = 2 \pi x_0 n_0 / \lambda$, and $\theta_s$ is a
discrete step over angles (in radians) which should be chosen
sufficiently small. We consider a Gaussian angular power
distribution with a width $\theta_0$,
\begin{equation}
G(\theta) = (\sqrt{2}/\sqrt{\pi} \theta_0) \exp \left(-
2\theta^2/\theta_0^2\right),
\end{equation}
so that the beams are fully coherent for $\theta_0 \rightarrow 0$,
and become less coherent as $\theta_0$ is increased.

We study a mutual interaction between solitons which are excited
by two beams launched in parallel,
\[
   E_s(x) = A \left\{ {\rm sech} [(x+d)/W_0] + e^{i\phi} {\rm sech}
[(x-d)/W_0] \right\} ,
\]
where $2 d$ is the separation between the beams at the input,
$W_0$ defines the beam widths, $\phi$ is the relative phase
between the two beams, and $A = \sqrt{2D/ \gamma W_0^2}$ is a
coherent soliton amplitude.

It is known that two equal-amplitude interacting coherent solitons
can experience attraction, repulsion, or energy exchange, and all
such processes are controlled by a single parameter, the soliton
relative phase $\phi$. As we show below, the interaction of
partially coherent beams depends critically on the coherence
parameter $\theta_0$.

\pict{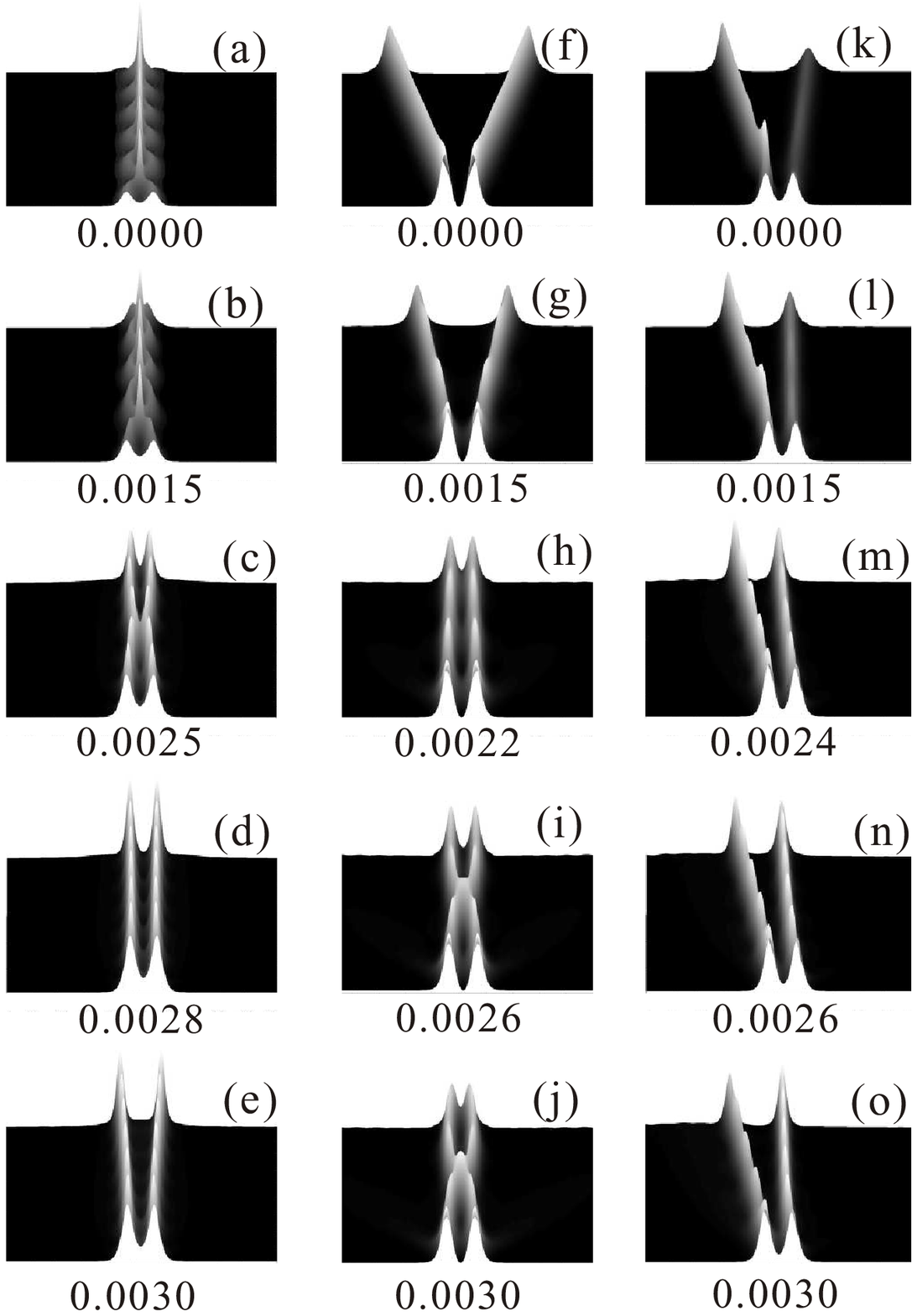}{fig1}{Numerical results. (a)-(e) Interaction of
in-phase solitons ($\phi=0$) for different values of the coherence
parameter $\theta_0$. (f)-(j) Interaction of $\pi$-out-of-phase
solitons ($\phi=\pi$). (k)-(o) Interaction of $\pi/2$-out-of-phase
solitons ($\phi=\pi/2$). }

In the simulations, which are performed to supplement the
experimental results presented below, we set
$\lambda=532$nm, $n_0=2.35$, $W_0=5 \mu$m, $2d=20 \mu$m, and the
total propagation length of 25 mm. For the in-phase interaction
($\phi=0$), Fig.~\rpict{fig1}(a) shows that the attraction between
two solitons makes the solitons launched in parallel cross each
other periodically when the entire field is {\em coherent}
($\theta_0=0$). We then increase the coherence parameter
$\theta_0$ from zero, i.e. make the entire field {\em partially
incoherent}. Since this raises the power threshold for a soliton
formation, we compensate for this effect by increasing the
strength of the nonlinearity $\gamma$. Our simulation results show
that the crossing period gets longer as the field becomes more
incoherent. When $\theta_0$ reaches the threshold value of 0.0028,
the solitons become propagating in parallel [Fig.~\rpict{fig1}(d)]
and, finally, when $\theta_0$ is larger than the threshold, the
solitons apparently {\em repel each other} [Fig.~\rpict{fig1}(e)].

For the $\pi$-out-of-phase solitons ($\phi=\pi$), the situation is
just opposite. Figures~\rpict{fig1}(f-j) show that the repulsion
between two solitons can be changed to attraction when the
incoherence parameter $\theta_0$ is larger than 0.0022. Also for the
$\pi/2$-out-of-phase interaction ($\phi=\pi/2$), after the
interaction, the situation that the right soliton is a bit
brighter than the left soliton due to energy transfer when the
entire field is coherent ($\theta_0=0$) can become opposite when
$\theta_0$ is above a threshold value around 0.0026
[Fig.~\rpict{fig1}(k-o)], though the opposite energy transfer is
not that obvious.

As the next step, we provide a theoretical explanation on how the
degree of coherence ($\theta_0$) can control the soliton
interaction. We note that due to the symmetry properties of
Eqs.~\reqt{nls}, the set of components $E_n$ can be chosen in a
number of different ways which provide exactly equivalent
descriptions~\cite{Soljacic:2003-254102:PRL}. In particular, it is possible to separate out symmetric and anti-symmetric contributions by making a unitary transformation
\[
\widetilde{E}_n = (E_{-n}+E_{n}) / \sqrt{2} = \cos( n \theta_s x )
E_s(x) \sqrt{2 G(n \theta_s) \theta_s}, \]
for $n>0$, $\widetilde{E}_0 = E_0 = E_s(x) \sqrt{G(0)
\theta_s}$, and
\[
\widetilde{E}_n = i (E_{-n}-E_{n}) / \sqrt{2} = \sin( n \theta_s x )
E_s(x) \sqrt{2 G(n \theta_s) \theta_s}, \]
for $n<0$. For a pair of in-phase solitons ($\phi=0$), the
components are symmetric, for $n\ge 0$, and they are
anti-symmetric, for $n < 0$; the situation is reversed for a pair
of out-of-phase solitons ($\phi=\pi$). This symmetry is preserved
during the propagation, and therefore there is no net energy
exchange between the in-phase or out-of-phase solitons. Thus, the
overall effect of attraction or repulsion between the solitons
depends on a balance between the powers of symmetric ($P_{\rm s}$)
and anti-symmetric ($P_{\rm as}$) components. A specific power
ratio $p = P_{\rm s} / P_{\rm as}$ is required to observe a
stationary propagation of two solitons in parallel, and it can be
estimated using an exact two-soliton solution~\cite{Kanna:2001-5043:PRL} of Eqs.~\reqt{nls}
as
\begin{equation}
p_0 \simeq 1 + 4 \exp( - 2 d / W_0 )
\end{equation}
for $2 d \gg W_0$.
At the crystal input, the parameter $p$
monotonously approaches the value of $+1$ from above (below) for a
pair of in-phase (out-of-phase) solitons as the incoherence
parameter $\theta_0$ is increased. However, since the
partially-coherent input field does not exactly match the soliton
profile, some energy is radiated away, and we found that the
parameter $p$ calculated for the spatially localized modes may
slightly increase or decrease during the initial propagation
stage. Because of this radiation mechanism, the condition $p=p_0$
can be satisfied at some threshold value of $\theta_0^{(\rm th)}$
when the solitons propagate in parallel.

We note that the strongest contribution to the $\cos$-type soliton
components ($n \ge 0$) comes from the angles around zero, whereas
the $\sin$-type components ($n<0$) are predominantly excited when
$n \theta_s \simeq \lambda / (4 d)$. Therefore, the soliton
separation $2d$ should be roughly inversely proportional to the
threshold value $\theta_0^{(\rm th)}$. We have performed
additional simulations to verify this prediction and found that
when the soliton separation $2 d$ is increased from 20 $\mu$m to 24
$\mu$m, the threshold value of the coherence parameter is indeed
reduced from 0.0028 to 0.0022, for the interaction of in-phase
solitons, and from 0.0022 to 0.0018, for the $\pi$-out-of-phase
solitons.

\pict{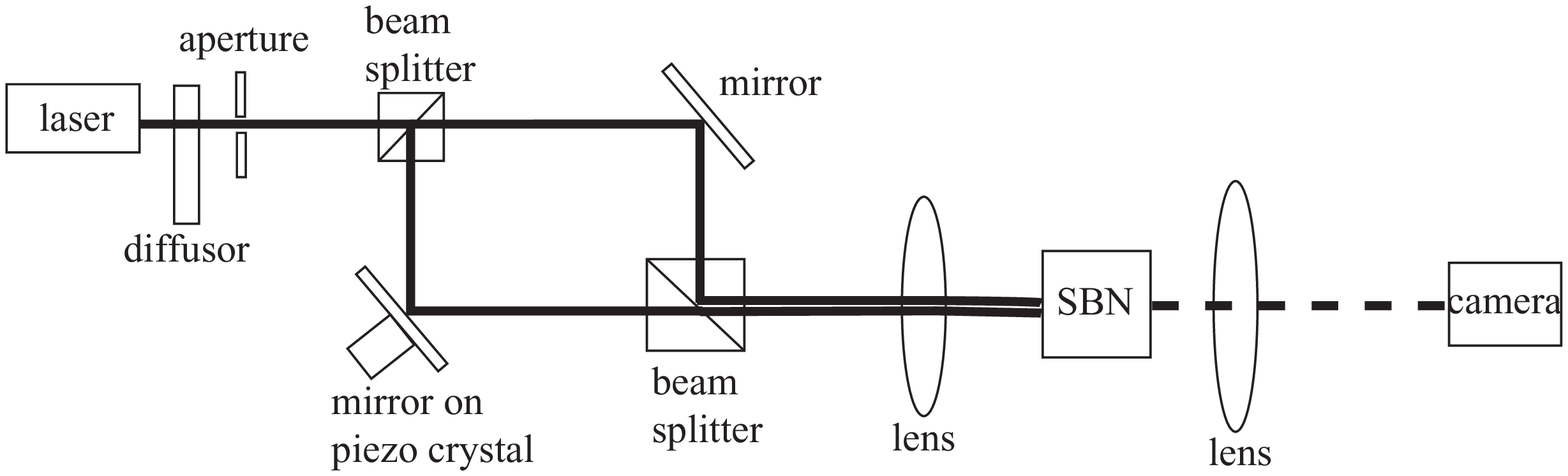}{fig3}{Sketch of the experimental setup. }

The interaction of two solitons with the initial $\phi=\pi/2$
phase shift is more complex. In addition to the effects of attraction
or repulsion, the solitons can exchange energy. At the initial
propagation stage, the energy flows from the right to the left
soliton in components $\widetilde{E}_n$ with $n\ge0$, but the flow
is opposite for $n<0$ because the phase difference is effectively
changed in the corresponding components from $\pi/2$ to $-\pi/2$.
As the incoherence is increased, the combined powers of two types
of components become almost equal, and overall energy exchange is
largely suppressed as observed in simulations.

\pict{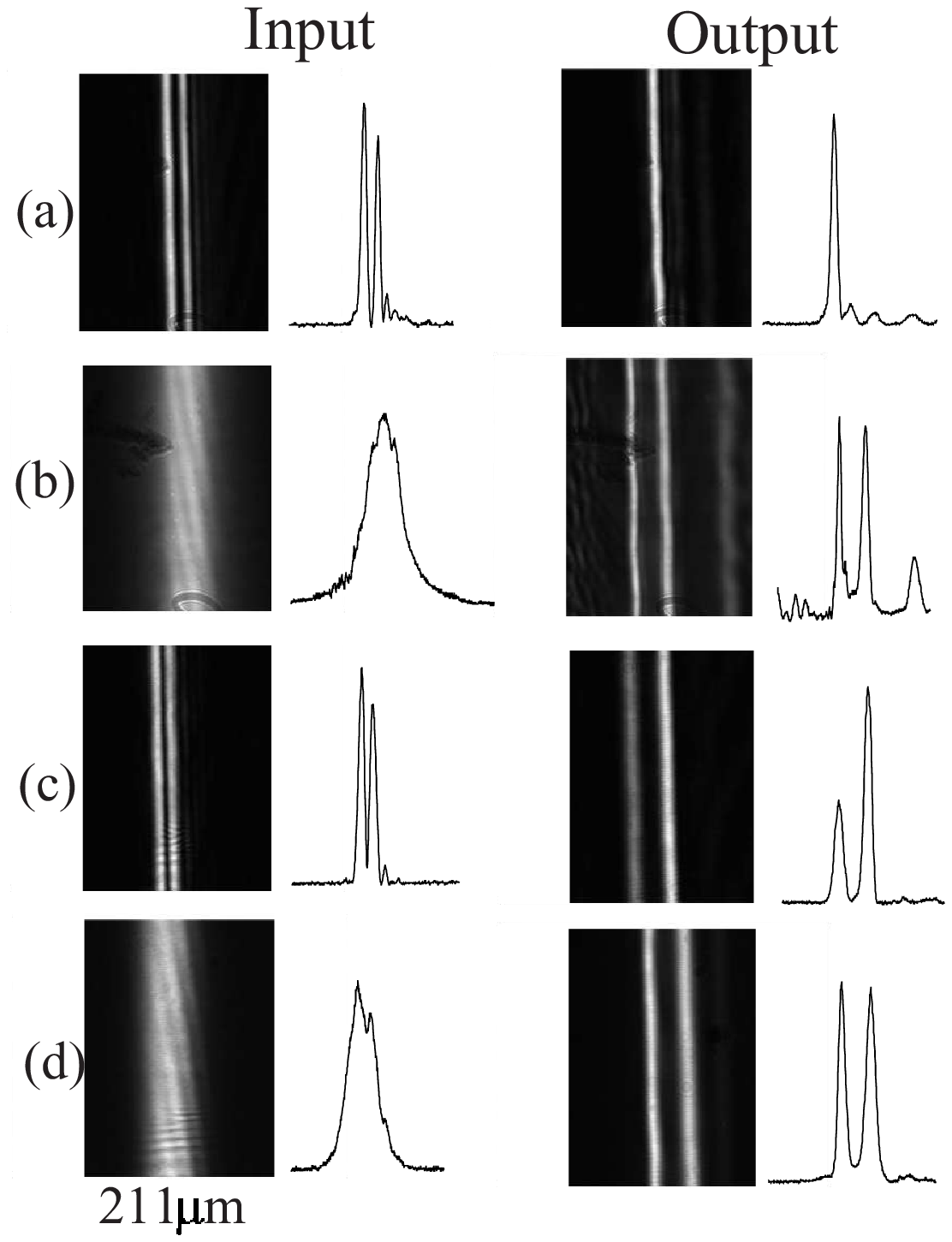}{fig4}{Experimental results for the interaction of
a pair of in-phase solitons for (a) coherent light and (b)
incoherent light, and a pair of $\pi/2$-out-of-phase solitons for
(c) coherent light and (d) incoherent light.}

In our experiment, we use the interaction of stripe
photorefractive solitons~\cite{Meng:1997-448:OL} as the study
object. As schematically shown in Fig.~\rpict{fig3}, we pass the laser
beam through a rotating diffuser to make the light partially
incoherent. The light beam is then split into two parts. We then
made these two parts to propagate in parallel by using a second
beam splitter. These two parts are then focused one-dimensionally
by a cylindrical lens onto the front face of a biased
photorefractive crystal. The entire crystal is illuminated with a
background intensity that is a bit stronger than the soliton
intensity to make the nonlinearity close to the Kerr-type
nonlinearity~\cite{Segev:1994-3211:PRL} but not much stronger to
avoid the transverse instability~\cite{Mamaev:1996-870:PRA}.

\pict{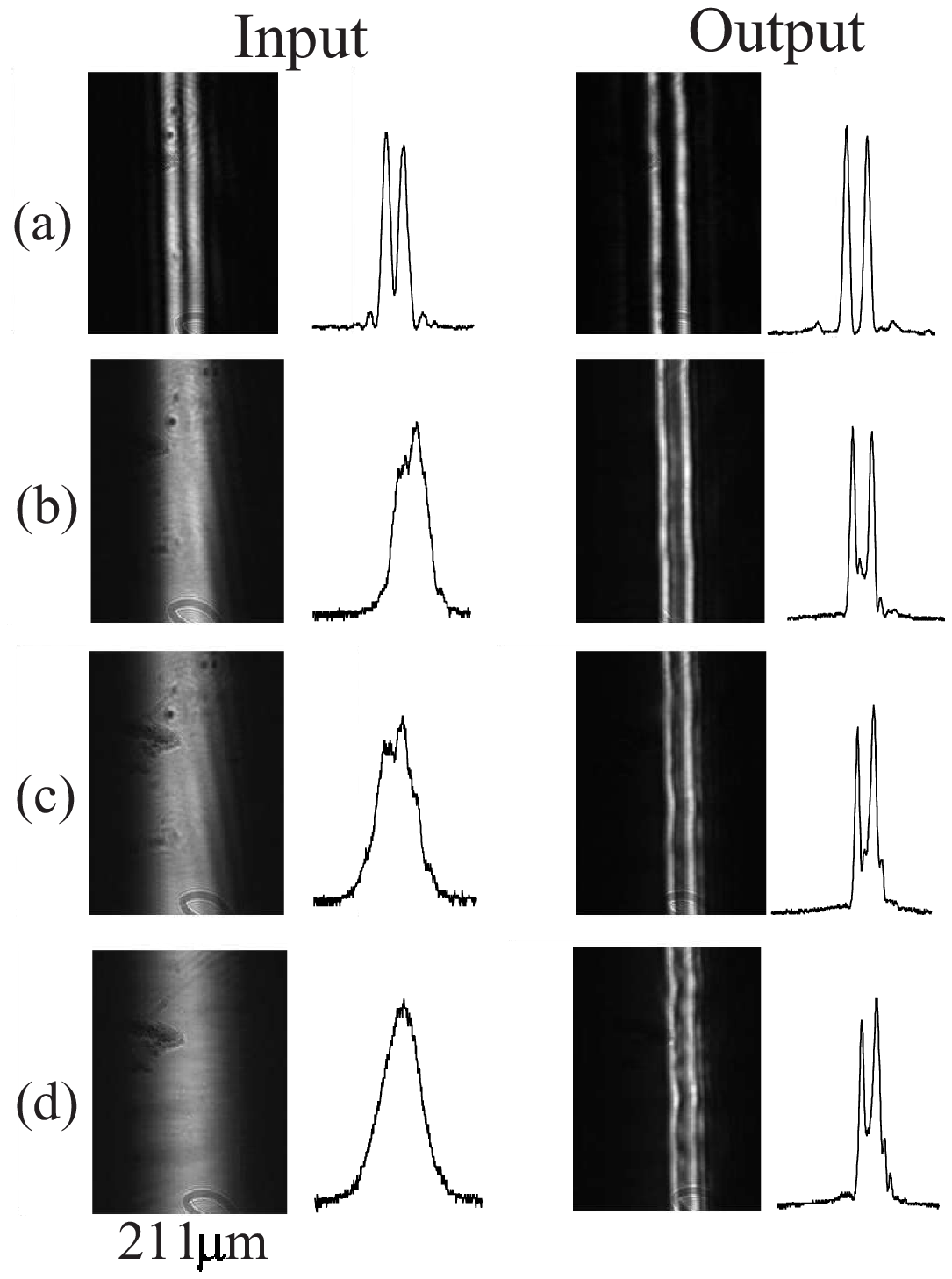}{fig5}{Experimental results for the interaction of
a pair of $\pi$-out-of-phase solitons. The soliton separation at
the output face decreases from the most coherent light, in the
case (a), to the least coherent light, in the case (d).}

We first take out the rotating diffuser to make the entire field
coherent. We also tune the optical path by applying a suitable
voltage on the piezo-transducer attached to the back of one
turning mirror to make the two parts in-phase. We then observe,
due to the in-phase attraction, these two parts merge into one at
the output face of the crystal [Fig.~\rpict{fig4}(a)]. We keep the
optical paths fixed for the entire system, but now insert the
rotating diffuser to make the total field partially incoherent. We
also increase the nonlinearity correspondingly, by higher biasing
voltage, due to the fact that incoherent solitons need higher
nonlinearity~\cite{Mitchell:1996-490:PRL}. As we can see at the
output face of the crystal [Fig.~\rpict{fig4}(b)], these two parts
apparently repel each other. This indeed confirms our simulation
that if the entire field is made incoherent enough, the attractive
interaction can change to a repulsive one. We then tune the
relative phase to be $\pi/2$. When the entire field is coherent,
there is strong energy transfer from the left to the right
solitons [Fig.~\rpict{fig4}(c)]. However, when the entire field is
partially incoherent enough, we observe the left one is a bit
brighter than the right one [Fig.~\rpict{fig4}(d)], very similar
to the simulation results shown in Fig.~\rpict{fig1}(o).

Finally, we  set the relative phase to be $\pi$. We adjust the
position of the diffuser and the size of the aperture to make the
field to be as coherent as possible. Figure~\rpict{fig5}(a) shows
the repulsion between the two solitons at the output face of the
crystal. We then increase the incoherence by opening the aperture
at a larger size. As the incoherence is increased, the separation
become smaller and can be smaller than the separation at the input
face, indicating the repulsion have been changed to attraction.

In conclusion, we have described theoretically and demonstrated
experimentally a new type of soliton interaction which is observed
when two solitons {\em as a whole} are made partially incoherent.
The interaction strength and type, either attraction or repulsion
between the solitons, can be controlled by varying their total
coherence. Even more, the interaction may change from attractive
to repulsive near a certain threshold in the coherence parameter.
We believe this novel type of the coherence controlled soliton
interaction is generic for the interaction of partially coherent
waves in nonlinear media, and it can be found in other fields.

This work was supported by a collaborative program between the
Australian Academy of Science and the National Science Council of
Taiwan, and by the Australian Research Council. One of the authors
(Yu.K.) thanks the Physics Department of the National Taiwan
University for hospitality.

\end{sloppy}
\end{document}